\documentstyle[prl,aps,epsfig,floats]{revtex}

\begin{document}
\draft
\wideabs{
  
  \title{Theory of angular magnetoresistance oscillations in $\bf
    Tl_2Ba_2CuO_6$}
  
  \author{Adrian Dr\u{a}gulescu and Victor M.~Yakovenko}
  \address{Department of Physics and Center for Superconductivity
    Research, University of Maryland, College Park, Maryland 20742}
  \author{David J.~Singh} \address{Complex Systems Theory Branch,
    Naval Research Laboratory, Washington DC 20375}

\date{\bf cond-mat/9811101, v.1: November 6, 1998, v.2: March 28, 1999}
\maketitle

\begin{abstract}Using the calculated electron energy band structure of
  $\rm Tl_2Ba_2CuO_6$, we compute the dependence of the {\bf c}
  axis magnetoresistance on the orientation of the magnetic field for
  different magnitudes of the magnetic field.  We explain the known
  experimental results for the in-plane rotation of the magnetic field
  and predict the shape of the magnetoresistance oscillations for the
  out-of-plane rotations of the magnetic field.  We show how the latter
  oscillations can be utilized to reconstruct the shape of the Fermi
  surface and to study the coherence of interplane electron
  tunneling.
\end{abstract}
}

In a strong magnetic field, the electrical resistivity of a layered
metal oscillates when the magnetic field is rotated between the
orientations perpendicular and parallel to the layers.  This effect,
called the angular magnetoresistance oscillations (AMRO), was
originally discovered in the layered organic conductors of the
BEDT-TTF family (see review \cite{Kartsovnik96}) and remains largely
unknown outside of the organic conductors research community.  AMRO
should not be confused with the Shubnikov-de Haas magnetoresistance
oscillations, which occur when the \emph{magnitude}, not the
\emph{orientation}, of the magnetic field is changed.  Using AMRO it
is possible to determine not only the area, but also the shape of the
Fermi surface of a layered metal \cite{Yakovenko92a}.  It would be
very interesting to observe AMRO experimentally in the high-$T_c$
superconductors, which are also layered materials.  That would provide
information about the structure of their Fermi surfaces and the
coherence of interlayer electron motion.

In this paper, we calculate the dependence of the {\bf c}-axis
(interlayer) electrical resistivity of $\mathrm{Tl_2Ba_2CuO_6}$ on the
orientation of a magnetic field and obtain detailed AMRO curves for
comparison with the past and future experiments \cite{BSCO}.  The
greater the value of $\omega_c\tau$, where $\omega_c$ is the cyclotron
frequency and $\tau$ is the electron scattering time, the more
pronounced AMRO are.  We believe that $\mathrm{Tl_2Ba_2CuO_6}$ is one
of the best candidates for the experimental observation of AMRO
because a relatively high value $\omega_c\tau=0.9$ has been recently
achieved in this material using a pulsed magnetic field of 60 T
\cite{Boebinger98}.  Moreover, AMRO for the magnetic field rotation
within the most conducting $({\bf a},{\bf b})$ plane have been already
observed experimentally in this material at the field of 13 T and
$\omega_c\tau=0.16\div0.31$ \cite{Cooper96}.

$\mathrm{Tl_2Ba_2CuO_6}$ has the body-centered tetragonal crystal
structure with in-plane lattice spacings $a=b=3.6$ {\AA} and the
distance $d=11$ {\AA} between the $\rm CuO_2$ planes.  The
electron band structure of $\mathrm{Tl_2Ba_2CuO_6}$ has been
calculated in Ref.\ \cite{Singh92}.  Two energy bands cross the Fermi
level: the Cu-O hole band centered at the Brillouin zone corner X and
the Tl-O electron band centered at $\Gamma$.  The Tl-O Fermi surface
is a closed spheroid, which does not contribute significantly to the
conductivity, so it will be ignored in our calculations.  The Cu-O
band is the generic band of the layered high-$T_c$ cuprates.  Its
Fermi surface has the shape of a slightly corrugated cylinder, open
along the {\bf c} direction.  The electron dispersion law is a sum of
the in-plane and interplane terms:
\begin{equation} \label{energy}
  \epsilon(k_x,k_y,k_z)=\epsilon_\|(k_x,k_y)+\epsilon_\perp(k_x,k_y,k_z),
\end{equation}
where $\epsilon$ is the electron energy, and $k_x$, $k_y$, and $k_z$
are the electron wave vectors along the {\bf a}, {\bf b}, and {\bf c}
axes.  The in-plane dispersion law $\epsilon_\|(k_x,k_y)$ has been
calculated numerically in Ref.\ \cite{Singh92} on a mesh of
$32\times32$ points and interpolated in between.  For the interplane
dispersion law $\epsilon_\perp(k_x,k_y,k_z)$, we select the
tight-binding one, in accordance with the body-centered unit cell:
\begin{equation} \label{eperp}
  \epsilon_\perp=-8 t_\perp \cos(k_x a/2) \cos(k_y b/2) \cos(k_z d),
\end{equation}
where the interplane electron tunneling amplitude $t_\perp$ is much
smaller than the Fermi energy $E_F$.

The orientation of the magnetic field {\bf H} is characterized by the
polar angle $\theta$ relative to the {\bf c} ($z$) axis perpendicular
to the layers and the azimuthal angle $\phi$ relative to the {\bf a}
($x$) axis parallel to the Cu-O bonds.  Within a semiclassical
picture, the electron wave vector {\bf k} changes in time $t$
according to the Lorentz equation of motion:
\begin{equation} \label{eq:Lorentz}
  d{\bf k}/dt= (e/\hbar c)\,{\bf v}\times{\bf H},
  \quad{\rm where}\quad
  {\bf v}=\partial\epsilon/\hbar\partial{\bf k}.
\end{equation}
Here $e$ is the electron charge, $c$ is the speed of light, and
$\hbar$ is the Planck constant.

The interplane component of the electrical conductivity tensor
$\sigma_{zz}$, obtained by solving the linearized Boltzmann equation
in the $\tau$ approximation, is given the Shockley-Chambers formula
\cite{Ziman}:
\begin{eqnarray} \label{sigma3}
\sigma_{zz}=2 e^2 \int\frac{d^3 k(0)}{(2\pi)^3}
     \left(-\frac{\partial f}{\partial\epsilon}\right)v_z[{\bf k}(0)]
\nonumber \\
     \times \int_0^{\infty} dt\, v_z[{\bf k}(t)]\, e^{-t/\tau}.
\end{eqnarray}
In Eq.\ (\ref{sigma3}), $f$ is the Fermi distribution function, $\tau$
is the electron scattering time, and the first integral is taken over
the electron wave vector ${\bf k}(0)$ that serves as the initial
condition for determining ${\bf k}(t)$ from the Lorentz equation of
motion (\ref{eq:Lorentz}).

When $\theta\neq0$, electrons circle around the Fermi surface along
the closed orbits obtained by cutting the Fermi surface with the
planes perpendicular to the magnetic field.  The different planes are
labeled by the component $k_H$ of the electron wave vector parallel to
the magnetic field.  In this case, Eq.\ (\ref{sigma3}) can be
rewritten as
\begin{eqnarray} \label{sigma1}
  &&\sigma_{zz}=\frac{e^3 H}{4 \pi^3 \hbar^2 c}\,
  \frac{1}{1-\exp{(-T/\tau)}}
\\ \nonumber 
  &&\times\int dk_H \int_{0}^{T}dt\,v_z[{\bf k}(t)] 
  \int_{0}^{T}dt'\,v_z[{\bf k}(t-t')]\,e^{-t'/\tau},
\end{eqnarray}
where $T$ is the period of electron motion.

\begin{figure}
\centerline{\psfig{file=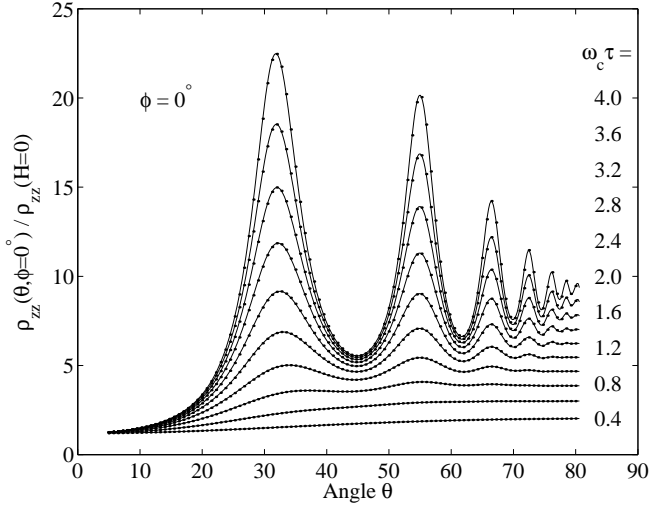,width=\linewidth,angle=0,clip=}}
%\centerline{\psfig{file=rotlfi0.eps,width=\linewidth,angle=0,clip=}}
\caption{ Angular oscillations of $\rho_{zz}$ vs $\theta$
  for $\phi=0^\circ$ at different values of $\omega_c\tau$. }
\label{fig:phi0}
\end{figure}

Because $t_\perp/E_F$ is very small, we neglect $v_z$ in Eq.\ 
(\ref{eq:Lorentz}), which then reduces to the in-plane equations of
motion for $k_x(t)$ and $k_y(t)$, while $k_z(t)$ is determined from
the geometrical relation
\begin{equation} \label{kz}
   k_z(t) = K_z - k_\phi(t)\tan\theta.
\end{equation}
Here $K_z=k_H/\cos\theta$, and
\begin{equation}
   k_\phi(t)=k_x(t)\cos\phi+k_y(t)\sin\phi
\end{equation}
is the projection of the in-plane electron wave vector onto the
in-plane component of the magnetic field \cite{Yakovenko92a}.

Substituting $v_z$ from Eq.\ (\ref{eperp}) into Eq.\ (\ref{sigma1}),
using Eq.\ (\ref{kz}), and taking the integral over $K_z$, we find:
\begin{eqnarray} \label{sigma2}
  &&\sigma_{zz}(\theta,\phi)=\frac{16 e^2 m_c t_\perp^2 d}
  {\pi^2 \hbar^4 \omega_c \cos\theta}\,
  \frac{1}{1-\exp{(-2\pi/ \omega_c\tau\cos\theta)}} 
\nonumber \\ 
  &&\times\int_{0}^{2\pi} d\zeta\int_{0}^{2\pi}d\zeta'
  \cos\{d [k_\phi(\zeta)-k_\phi(\zeta-\zeta')]\tan\theta\} 
\nonumber\\ 
  &&\times \cos\frac{ak_x(\zeta)}{2}\,
  \cos\frac{ak_y(\zeta)}{2}\,
  \cos\frac{ak_x(\zeta-\zeta')}{2}\,
  \cos\frac{ak_y(\zeta-\zeta')}{2}
\nonumber \\
  &&\times \exp(-\zeta'/\omega_c\tau\cos\theta). 
\end{eqnarray}
In Eq.\ (\ref{sigma2}), $t$ has been replaced by the dimensionless
variable $\zeta=\omega_c t$, where $\omega_c$ is the in-plane
cyclotron frequency for $\theta=0$:
\begin{equation} \label{omega}
  \omega_c=\frac{2\pi e H}{c\hbar\oint dk_l/v} 
  = \frac{eH}{m_c c}.
\end{equation}  
In Eq.\ (\ref{omega}), the integral is taken along the Fermi surface
in the $(k_x,k_y)$ plane, and $m_c$ is, by definition, the cyclotron
mass.

\begin{figure}
\centerline{\psfig{file=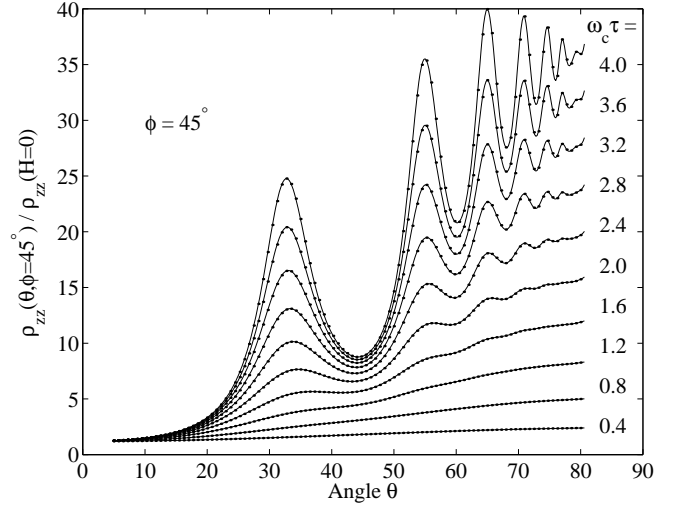,width=\linewidth,angle=0}}
%\centerline{\psfig{file=rotlfi45.eps,width=\linewidth,angle=0}}
\caption{ Angular oscillations of $\rho_{zz}$ vs $\theta$
  for $\phi=45^\circ$ at different values of $\omega_c\tau$. }
\label{fig:phi45}
\end{figure}

\begin{figure}
\centerline{\psfig{file=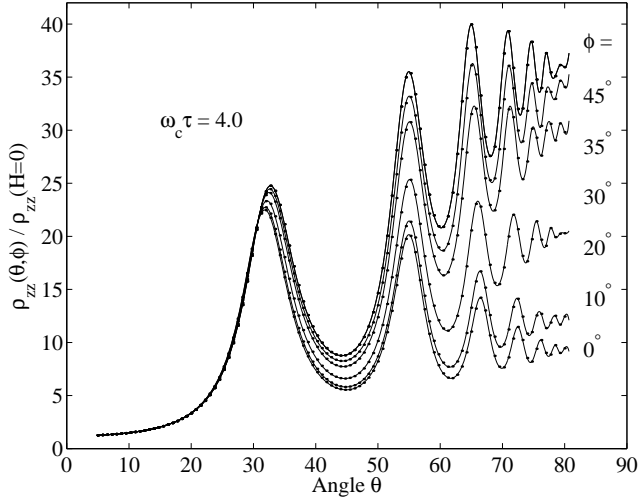,width=\linewidth,angle=0}}
%\centerline{\psfig{file=rotlallfi.eps,width=\linewidth,angle=0}}
\caption{ Angular oscillations of $\rho_{zz}$ vs $\theta$
  for different $\phi$ at $\omega_c\tau=4.0$. The curve for
  $\phi=40^\circ$ is the same as for $\phi=45^\circ$.}
\label{fig:phi_all}
\end{figure}

We have integrated the Lorentz equations of motion for $k_x(t)$ and
$k_y(t)$ and evaluated the integral (\ref{sigma2}) numerically for
different values of the parameter $\omega_c\tau$ and different
orientations of the magnetic field.  The resultant resistivity
$\rho_{zz}=1/\sigma_{zz}$ \cite{z_xy} is shown in Figs.\ 
\ref{fig:phi0}, \ref{fig:phi45}, and \ref{fig:phi_all}.  For a fixed
azimuthal angle $\phi$, $\rho_{zz}$ displays oscillations as a
function of the polar angle $\theta$.  Figs.\ \ref{fig:phi0} and
\ref{fig:phi45} demonstrate that, while the amplitude of oscillations
grows with increasing $\omega_c\tau$, the angles $\theta_n$ where
$\rho_{zz}$ has the $n$-th maximum do not depend on $\omega_c\tau$.

The values of $\theta_n$ can be related to the shape of the Fermi
surface via the following analytical argument \cite{Yakovenko92a}.
The maxima in $\rho_{zz}$ become sharp in the limit
$\omega_c\tau\gg1$.  In this limit, as follows from Eq.\ 
(\ref{sigma1}),
$\sigma_{zz}\propto\tau\langle\overline{v}_z^2\rangle$, where
$\langle\cdots\rangle$ denotes the averaging over different electron
orbits (the integration over $k_H$), and
\begin{eqnarray} \label{vz}
  \overline{v}_z=\frac{8 t_\perp d}{\hbar} \int_{0}^{2\pi}d\zeta
  \cos\frac{ak_x(\zeta)}{2}\,\cos\frac{ak_y(\zeta)}{2}
\nonumber \\
  \times\sin\left[K_zd - k_\phi(\zeta)\,d\,\tan\theta\right]
\end{eqnarray}  
is the average of $v_z$ along a given electron orbit.  In the limit
$(d/a)\tan\theta\gg1$, the argument of sine in Eq.\ (\ref{vz})
oscillates very fast as a function of $\zeta$.  Thus the integral is
dominated by the points where the phase is stationary.  These are the
turning points of the electron trajectory, where $k_\phi$ achieves the
maximal $k_{\phi}^{\rm max}$ and minimal $-k_{\phi}^{\rm max}$ values.
Evaluating the integral (\ref{vz}) asymptotically in the vicinity of
the stationary points, we find \cite{Yakovenko92a}:
\begin{equation} \label{vzbar}
  \overline{v}_z\propto\sin(K_zd)\,
  \cos(k_\phi^{\rm max}\,d\,\tan\theta -\pi/4).
\end{equation}
When the argument of cosine in Eq.\ (\ref{vzbar}) is equal to
$\pi(n-1/2)$ (where $n=1,2,\ldots$), the average electron velocity
vanishes: $\overline{v}_z=0$, thus $\sigma_{zz}\to0$ and
$\rho_{zz}\to\infty$ \cite{infinity}.  This situation corresponds to
the resistivity maxima in Figs.\ \ref{fig:phi0}, \ref{fig:phi45}, and
\ref{fig:phi_all}, and takes place at the angles
\begin{equation} \label{tangent}
  \tan{\theta_n}=\frac{\pi(n-1/4)}{k_\phi^{\rm max}d}.
\end{equation}
The same condition (\ref{tangent}) follows from the derivation even if
$\tau$ varies along the Fermi surface \cite{Yakovenko98a}.

\begin{figure}
\centerline{\psfig{file=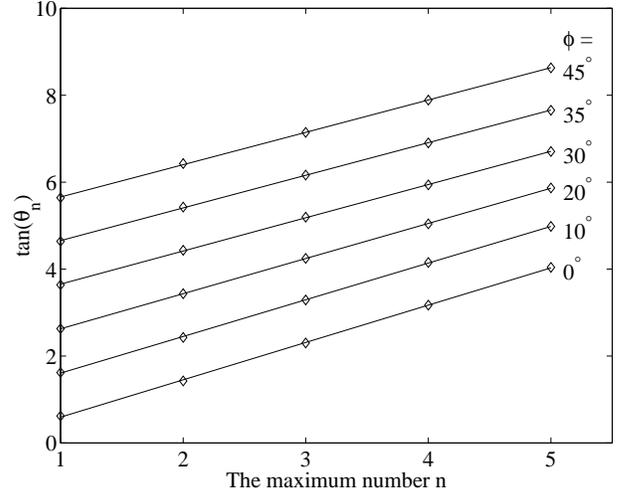,width=0.95\linewidth,angle=0,clip=}}
%\centerline{\psfig{file=tanmaxtl.eps,width=\linewidth,angle=0,clip=}}
\caption{ The linear dependence of $\tan\theta_n$ vs the maximum
  number $n$. The lines are offset vertically for different $\phi$.}
\label{fig:tanmax}
\end{figure}

\begin{figure}
\centerline{\psfig{file=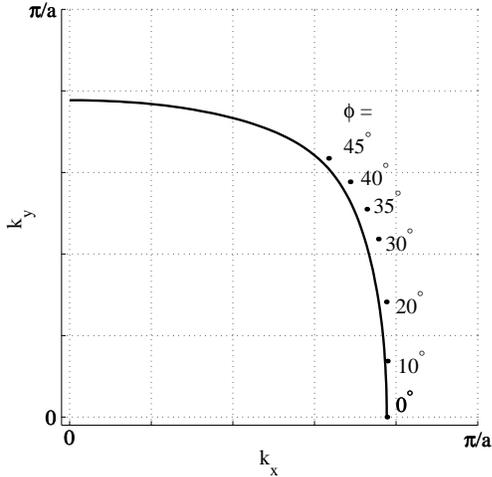,width=0.8\linewidth,angle=0,clip=}}
%\centerline{\psfig{file=recofs.eps,width=\linewidth,angle=0,clip=}}
\caption{ Dots: ${\bf k}_\phi^{\rm max}$ vs $\phi$.  
  Solid line: The Fermi surface of $\mathrm{Tl_2Ba_2CuO_6}$.  The
  normals to the radii defined by the dots and drawn through the dots
  envelope the Fermi surface. }
\label{fig:FS}
\end{figure}

Eq.\ (\ref{tangent}) shows that $\tan{\theta_n}$ increases linearly
with the maximum number $n$, and the slope of this dependence is
determined by $k_\phi^{\rm max}$.  In Fig.\ \ref{fig:phi_all}, we plot
$\rho_{zz}$ vs $\theta$ for different $\phi$.  For each curve, we
determine the angles of the resistivity maxima $\theta_n$ and plot
$\tan{\theta_n}$ vs $n$ in Fig.\ \ref{fig:tanmax}.  Using Eq.\ 
(\ref{tangent}), we determine $k_\phi^{\rm max}$ from the slopes of
the lines in Fig.\ \ref{fig:tanmax} and plot $k_\phi^{\rm max}$ vs
$\phi$ in Fig.\ \ref{fig:FS} as the black dots.  To reconstruct the
Fermi surface, it is necessary to draw a line through each black dot
in Fig.\ \ref{fig:FS} perpendicular to the vector ${\bf k}_\phi^{\rm
  max}$.  The envelope of these lines gives the Fermi surface
\cite{Yakovenko92a}.  One can check in Fig.\ \ref{fig:FS} that this
procedure indeed reproduces the actual Fermi surface of our model
\cite{Singh92}, which is shown by the solid line.  Thus, measuring
AMRO in $\mathrm{Tl_2Ba_2CuO_6}$ experimentally for every angle $\phi$
and applying the described above procedure, one could reconstruct the
in-plane shape of the Fermi surface in this material.

Now, let us discuss AMRO of $\rho_{zz}$ vs $\phi$ for the magnetic
field rotation within the $(x,y)$ plane at $\theta=90^\circ$.  In this
case, the electron orbits are open along the {\bf c} axis.  Neglecting
$v_z$ in Eq.\ (\ref{eq:Lorentz}), we find that only $k_z$ depends on
time: $k_z(t)=k_z(0)-(eH/c\hbar)\,(v_x\sin\phi-v_y\cos\phi)\,t$.
Neglecting $\epsilon_\perp$ in Eq.\ (\ref{energy}), we can write the
volume of integration in Eq.\ (\ref{sigma3}) as
$d^3k(0)=dk_z(0)\,d\epsilon_\|\,dk_l/\hbar v$, where the differential
$dk_l$ is taken along the Fermi surface in the $(k_x,k_y)$ plane.
Substituting $v_z$ from Eq.\ (\ref{eperp}) into Eq.\ (\ref{sigma3}),
using $k_z(t)$, and taking the integrals over $\epsilon_\|$, $k_z(0)$,
and $t$, we find:
\begin{equation} \label{sigma4}
  \sigma_{zz}(90^\circ,\phi)=\frac{8 e^2 t_\perp^2 \tau d}
  {\pi^2\hbar^3} \oint \frac{dk_l}{v}\,
  \frac{\cos^2\frac{ak_x}{2}\,\cos^2\frac{ak_y}{2}}
%  \frac{\cos^2(\tilde{k}_x/2)\,\cos^2(\tilde{k}_y/2)}
  {1+[\omega_z(\phi)\tau]^2}.
\end{equation}
In Eq.\ (\ref{sigma4}),
\begin{equation}  \label{eq:omega_z}
  \omega_z(\phi)=\frac{ed}{\hbar c}|{\bf v}\times{\bf H}|
  =\frac{edH}{\hbar c}|v_x\sin\phi-v_y\cos\phi|
\end{equation}
is the frequency of the electron motion across the Brillouin zone in
the $k_z$ direction.  Similar equations were obtained in Ref.\ 
\cite{Lebed97b}.

\begin{figure}
\centerline{\psfig{file=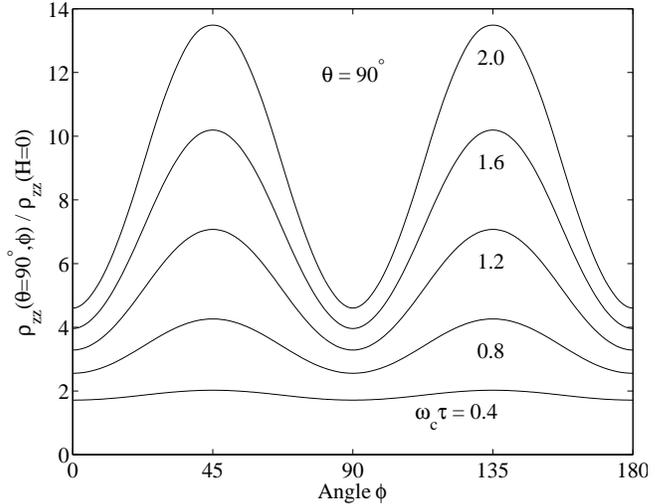,width=\linewidth,angle=0,clip=}}
%\centerline{\psfig{file=ro90tl.eps,width=\linewidth,angle=0,clip=}}
\caption{Angular oscillations of $\rho_{zz}$ vs $\phi$
  for $\theta=90^\circ$ at different values of $\omega_c\tau$.}
\label{fig:theta90}
\end{figure}

Using Eq.\ (\ref{sigma4}), we have numerically calculated
$\rho_{zz}(90^\circ,\phi)$ and plotted it in Fig.\ \ref{fig:theta90}
for different values of $\omega_c\tau$.  One can see that $\rho_{zz}$
is minimal at $\phi=0^\circ$ and maximal at $\phi=45^\circ$. This
result is in agreement with Fig.\ \ref{fig:phi_all} and with the
experiment \cite{Cooper96}.  This behavior can be qualitatively
understood in the following way.  According to Eqs.\ (\ref{sigma4})
and (\ref{eq:omega_z}), $\sigma_{zz}(90^\circ,\phi)$ is dominated by
the regions of the Fermi surface where $\omega_z(\phi)$ is minimal,
that is where {\bf v} is parallel to {\bf H} \cite{Lebed97b}.  When
the magnetic field points along the $x$ axis, these regions are
relatively flat and large (see Fig.\ \ref{fig:FS}).  On the other
hand, when the magnetic field points at $45^\circ$ to the $x$ axis,
these regions are curved and relatively small.  Thus, the conductivity
is higher when {\bf H} is parallel to $x$ \cite{Bulaevskii}.  

The experiment \cite{Cooper96} was actually performed in the regime
$\omega_c\tau\ll1$.  In this case, $\rho_{zz}$ can expanded in the
powers of the magnetic field $H$:
$\rho_{zz}=\rho_{zz}^{(0)}+\Delta\rho_{zz}^{(2)}-\Delta\rho_{zz}^{(4)}$,
where $\rho_{zz}^{(0)}$ is the zero-field resistance, and
$\Delta\rho_{zz}^{(2)}$ and $\Delta\rho_{zz}^{(4)}$ are the positive
terms proportional to $H^2$ and $H^4$, respectively.  Angular
dependence appears only in the $\Delta\rho_{zz}^{(4)}$ term, which can
be written as $\Delta\rho_{zz}^{(4)}=\bar\rho_{zz}^{(4)}
+\tilde\rho_{zz}^{(4)}\cos(4\phi)$. Expanding Eq.\ (\ref{sigma4}) in
the powers of $\omega_c\tau$ and calculating the integrals
numerically, we find the value 0.505 for the dimensionless ratio
$\bar\rho_{zz}^{(4)}\rho_{zz}^{(0)}/[\Delta\rho_{zz}^{(2)}]^2$, which
does not depend on the magnetic field and the scattering time $\tau$.
This value agrees with the experimentally measured one $0.6\pm 0.1$
\cite{Cooper96}.  We also obtain the value 0.16 for the dimensionless
ratio of the angular-dependent and angular-independent terms
$\tilde\rho_{zz}^{(4)}/\bar\rho_{zz}^{(4)}$.  The experiment
\cite{Cooper96} finds that this ratio varies with temperature between
0.15 and 0.06.  The temperature dependence may be due to different
temperature dependences of $\tau$ at the different parts of the Fermi
surface \cite{Yakovenko98a}, which is not considered in our model.

Our results are obtained from the Boltzmann equation in the lowest
order in the interplane electron tunneling amplitude $t_\perp$, which
appears only as a prefactor in Eqs.\ (\ref{sigma2}) and
(\ref{sigma4}).  Equivalent results can be obtained using the
lowest-order perturbation theory in $t_\perp$ and the in-plane
electron Green functions \cite{McKenzie}.  AMRO exist both when the
interplane electron motion is coherent $t_\perp\geq\hbar/\tau$ or
weakly incoherent $t_\perp\leq\hbar/\tau$ \cite{McKenzie}.  On the
other hand, if the interplane tunneling is strongly incoherent, so
that electrons loose phase memory, and the in-plane electron momentum
is not conserved, AMRO should not exist.  The experimental observation
of AMRO in $\mathrm{Tl_2Ba_2CuO_6}$ \cite{Cooper96} indicates that the
interplane tunneling in this material corresponds to the former case.

\vspace{-1.5\baselineskip}


\begin{references}
\bibitem{BSCO} We have also calculated AMRO for $\rm
  Bi_2Sr_2CaCu_2O_8$ using the electron dispersion law determined from
  the photoemission experiment \cite{Campuzano95} and found results
  similar to those for $\rm Tl_2Ba_2CuO_6$.
\bibitem{z_xy} The off-diagonal components of the conductivity tensor
  $\sigma_{zx}$ and $\sigma_{zy}$, given by equations similar to Eq.\ 
  (\ref{sigma1}), vanish because of the integration of $v_z$ over
  $K_z$, so $\rho_{zz}=1/\sigma_{zz}$.
\end{references}
\end{document}